\documentclass[preprint,showpacs,preprintnumbers,amsmath,amssymb]{revtex4}
\usepackage{graphicx}
\usepackage{dcolumn}
\usepackage{bm}
\usepackage{epsfig}

\newcommand{\geqnew}{\stackrel{>}{\!\ _{\sim}}}

\begin{document}

\preprint{PSU/TH-250}

\title{Quantum wave packet revivals in circular billiards}

\author{R. W. Robinett} \email{rick@phys.psu.edu}
\author{S. Heppelmann}\email{heppel@phys.psu.edu}
\affiliation{%
Department of Physics\\
The Pennsylvania State University\\
University Park, PA 16802 USA \\
}%

\date{\today}

\begin{abstract}

We examine the long-term time-dependence of Gaussian wave packets in a 
circular infinite well (billiard) system and find that there are 
 approximate revivals. For the special case of purely $m=0$ states 
(central wave packets with no momentum) the revival time is 
$T_{rev}^{(m=0)} = 8\mu R^2/\hbar \pi$, where $\mu$ is the mass of the 
particle, and the revivals are almost exact. For all other wave packets,
we find that $T_{rev}^{(m \neq 0)} = (\pi^2/2) T_{rev}^{(m=0)}
\approx 5T_{rev}^{(m=0)}$ and the nature of the revivals becomes 
increasingly approximate as the average angular momentum or number 
of $m \neq 0$ states  is/are increased. 
The dependence of the revival structure on the initial position, energy, 
and angular momentum of the wave packet and the connection to the energy 
spectrum is discussed in detail. The results are also compared to
two other highly symmetrical 2D infinite well  geometries with exact 
revivals, namely the square and equilateral triangle billiards. We also
show explicitly how the classical periodicity for closed orbits in a circular
billiard arises from the energy eigenvalue spectrum, using a WKB analysis.

\end{abstract}


\maketitle

\section{\label{sec:level1} Introduction}

The connections between the quantized energy eigenvalue spectrum of a bound
state and the classical motions of the corresponding classical
system have become increasingly  interesting and important as the ability 
to experimentally
probe the quantum-classical interface has dramatically improved. Methods
such as periodic orbit theory \cite{gutzillwer}, \cite{semiclassical},
for example, provide direct connections between the energy spectrum 
and the closed 
orbits of the classical system. The study of the time-dependence of
wave packet solutions also depends critically on the energy spectrum,
most especially as related to the existence of revivals (and superrevivals), 
wherein  initially localized states which have a short-term, quasi-classical 
time evolution, 
can spread significantly over several orbits, only to reform later in the 
form of a quantum revival in which the spreading reverses itself, the wave 
packet relocalizes, and the semi-classical periodicity is once again evident.
Such revival phenomena have been observed in a wide variety of physical
systems, especially in Rydberg atoms \cite{revival_review}, 
 and calculations exist for many other systems \cite{other_revivals}.

The  archetype of a one-dimensional model system for quantum  revivals
is the infinite well (where such revivals are exact) and a number
of analyses 
\cite{1d_fractional}
-- \cite{styer} 
have provided insight into both the short-term and long-term 
behavior of wave packets. 
Just as with many systems which depend on a single quantum
number, one typically expands the energy eigenvalues (assuming integral
values) about the central value used in the construction of a wave packet
via 
\begin{equation}
E(n) \approx E(n_0) + E'(n_0)(n-n_0) + \frac{1}{2}E''(n_0)(n-n_0)^2
+ \frac{1}{6}E'''(n_0)(n-n_0)^3 + \cdots
\end{equation}
in terms of which the classical period, revival, and superrevival times
are given respectively by
\begin{equation}
T_{cl} = \frac{2\pi\hbar}{|E'(n_0)|}
\qquad \quad
T_{rev} = \frac{2\pi \hbar}{|E''(n_0)|/2}
\qquad \quad
T_{super} = \frac{2\pi \hbar}{|E'''(n_0)|/6}
\end{equation}

Systems with two quantum numbers \cite{bluhm_2d}, \cite{other_2d},
with energies labeled by $E(n_1,n_2)$,
offer richer possibilities for wave packet revivals and typically the 
long-term revival structure depends on three possible times, given by
\begin{equation}
T_{rev}^{(n_1)} = \frac{2\pi \hbar}{(1/2)\partial^2 E(n_1,n_2)/\partial n_1^2}
\quad
\quad
T_{rev}^{(n_2)} = \frac{2\pi \hbar}{(1/2)\partial^2 E(n_1,n_2)/\partial n_2^2}
\end{equation}
and
\begin{equation}
T_{rev}^{(n_1,n_2)} = \frac{2\pi \hbar}{\partial^2 E(n_1,n_2)/\partial n_1 \partial n_2}
\end{equation}
and the revival structure depends on the interplay between these three times.

The two-dimensional generalization of the infinite well, or
two-dimensional infinite square well (billiard) of size $L \times L$, 
provides the simplest example of such a two quantum number system,  
and in this case the revival times are identical, namely
\begin{equation}
T_{rev}^{(n_x)} = \frac{4 m L^2}{\hbar \pi} = T_{rev}^{(n_y)}
\label{2d_revival_time}
\end{equation}
with no cross-term present. For rectangular infinite wells with incommensurate
sides ($L_x \times L_y$, $L_x/L_y \neq p/q$), the structure of the revival 
times may be more complex \cite{bluhm_2d}, \cite{other_2d}.

Another trivially related 2D billiard system for which exact quantum revivals
are guaranteed is an isosceles triangle infinite well with 
$45^{\circ}$ angles. 
The solutions for that system can be obtained from the 2D square well 
(cut along the diagonal) by 
taking appropriate linear combinations of the form
$[u_{(n_x)}(x)u_{(n_y)}(y) - u_{(n_x)}(y)u_{(n_y)}(x)]/\sqrt{2}$ 
which then  satisfy
the boundary conditions along the long side (hypotenuse) of the triangle, 
so long as $n_x \neq n_y$.  The energy spectrum therefore consists of one 
set of the 2D square well energies, 
$E(n_x,n_y) = \hbar^2 \pi^2(n_x^2+n_y^2)/2mL^2$, 
but restricted to $n_x< n_y$. The purely quadratic energy dependence 
guarantees that all wave packets will have the standard revival time given by
Eqn.~(\ref{2d_revival_time}).

A less obvious case of exact quantum revivals in a novel  2D billiard shape
exists for the equilateral triangle. The energy spectrum for this system 
\cite{berry}, \cite{math_methods} (with side of length $L$) is  given by 
\begin{equation}
E(p,q) = \left(\frac{4}{3}\right)^2 
\left(\frac{\hbar^2 \pi^2}{2mL^2}\right)
(p^2 + q^2 - qp)
\qquad
\mbox{with $1\leq q \leq p/2$}
\label{triangle_energies}
\end{equation}
with all states being degenerate (due to symmetries about obvious axes),
except for those with $p=2q$. Because of the exact quadratic dependence
(with trivially related coefficients) on the two quantum numbers, there
are exact revivals with a common revival time given by 
\begin{equation}
T_{rev}^{(p)}
=
T_{rev}^{(q)}
=
T_{rev}^{(p,q)}
= \frac{9mL^2}{4\hbar \pi}
\label{triangle_revival_time}
\end{equation}

Given the fact that these two simple geometrical cases of 2D 
infinite well (billiard) systems exhibit exact quantum revivals, it
is of interest to study to what extent, if any, the circular infinite well
will exhibit revival structures. Just as  square/rectangular 
\cite{gutzwiller_paper} and spherical/circular \cite{balian_and_bloch} 
billiard geometries were among the first considered using 
periodic orbit theory, a comparison of the revival structure 
in these two distinct systems is appropriate.
While qualitatively different from the  equilateral triangle or square wells, 
one can perhaps suggestively describe 
these geometries as regular $N$-sided polygons  with $N=3$ and $N=4$
respectively, so that the circular case we will consider here
corresponds to $N= \infty$.

The ability to fabricate such billiard systems (or analogs thereof) and 
experimentally probe the energy spectrum or time-dependence makes such a 
study of more than academic interest. For example, the energy-level structure 
and statistics of microwave cavities \cite{microwave} provide an analog system
to probe periodic orbit theory and statistical measures of chaotic behavior
in arbitrary shaped 2D billiard  geometries. Measurements of conductance
fluctuations in ballistic microstructures \cite{microstructures} have
been tentatively used to identify frequency features in the power spectrum 
with 
specific closed orbits in a circular (and stadium) billiard. More recently, 
the realization of atom-optics billiards \cite{atom_optics}, with ultra-cold
atoms in arbitrary shaped 2D boundaries confined by optical dipole 
potentials,  has allowed the study of various chaotic and integrable shapes
such as the stadium, ellipse and circle, again for short-term, semi-classical
propagation. Motivated by these studies, we will focus on the long-term 
revival structure of quantum wave packets in a circular billiard geometry.

Since the revival structure of any system depends on the quantum
number dependence of the quantized energy eigenvalues, in Sec.~II we
discuss,  in some detail,  the dependence of the two-dimensional 
circular infinite well eigenvalues on the radial and angular momentum 
quantum numbers, $n_r$ and $m$.  
We then, in Sec.~III,  present numerical results for the autocorrelation
function for several classes of initially Gaussian wave packets
in the circular well. 

Based on both the  energy eigenvalue analysis and our detailed wave packet simulations, for purely $m=0$ wave packets (implying radially 
symmetric states, with no  average momentum) we find almost exact 
revivals with 
$T_{rev}^{(m=0)} = 8 \mu R^2/\hbar \pi$ where $\mu$ denotes the mass of the
point particle in the circular billiard of radius $R$. For wave packets
including $|m| \neq 0$ components, meaning any with non-vanishing average
momentum or not initially localized at the center of the well, 
there are only approximate revivals,
becoming increasingly so as the angular momentum is increased. The
revival times in all these cases is $T_{rev}^{(m\neq 0)} = 
(\pi^2/2)T_{rev}^{(m=0)} \approx 5 T_{rev}^{(m=0)}$ 
due to the seemingly accidental fact that $10/\pi^2 = 1.013$.

Finally, in one Appendix (A) we point out a similarity between a pattern of 
special 'accidentally'  revival times in the 2D circular, square, and 
equilateral triangle  wells, while in a second Appendix (B), 
 we show explicitly how the
classical periodicity for closed orbits in the circular well arises 
directly from the quantum mechanical energy spectrum, using a WKB analysis.

\section{ Energy spectrum for the circular infinite well }

The problem of a point particle (with mass denoted by $\mu$, to avoid confusion
with the angular momentum quantum number, $m$) confined to  a circular infinite
well of radius $R$ is defined by the potential
\begin{equation}
      V_C(r) = \left\{ \begin{array}{ll}
               0 & \mbox{for $r<R$} \\
               \infty  & \mbox{for $r\geq R$}
                                \end{array}
\right.
\end{equation}
The (unnormalized) solutions of the corresponding 2D Schr\"odinger equation 
are given by
\begin{equation}
\psi(r,\theta) = J_{|m|}(kr) e^{im\theta}
\end{equation}
where the quantized angular momentum values are given by $L_z = m \hbar$
for $m=0, \pm 1, \pm 2,...$ and  the $J_{|m|}(kr)$ are the (regular) 
Bessel functions of order $|m|$. 

The wavenumber, $k$,  is related to the  energy via 
$k = \sqrt{2\mu E/\hbar^2}$ and the energy eigenvalues are
quantized by application of the boundary conditions at the infinite wall
at $r=R$, namely $J_{|m|}(z=kR) =0$. The quantized energies are then given by
\begin{equation}
E_{(m,n_r)} = \frac{\hbar^2 [z_{(m,n_r)}]^2}{2\mu R^2}
\end{equation}
where $z_{(m,n_r)}$ denotes the zeros of the Bessel function of order
$|m|$,  and $n_r$ counts the number of radial nodes. The energy spectrum
is doubly degenerate for $|m| \neq 0$ corresponding to the equivalence
of clockwise and counter-clockwise ($m>0$ and $m<0$) motion. 
Because the quantum number dependence  of the energy eigenvalues is the
determining factor in the structure of wave packet revivals, we wish to
examine  the $m,n_r$ dependence of the $E_{(m,n_r)} \propto [z_{(m,n_r)}]^2$ 
in detail. 

As a first approximation, one can look at the large $z$ behavior of the 
Bessel function solutions \cite{math_handbook} for fixed values of $|m|$, 
namely
\begin{equation}
J_{|m|}(z) 
\longrightarrow 
\sqrt{\frac{2}{\pi z}}
\cos\left(z - \frac{|m|}{2}- \frac{\pi}{4}\right) + \cdots
\end{equation}
With this approximation, the zeros would be given by
\begin{equation}
z_{(m,n_r)} -\frac{|m|}{2} - \frac{\pi}{4} \approx  (2n_r+1)\frac{\pi}{2}
\qquad
\mbox{or}
\qquad
z_{(m,n_r)} \approx  \left(n_r+\frac{|m|}{2}+\frac{3}{4}\right)\pi \equiv
z_{0}(m,n_r)
\label{first_approximation}
\end{equation}
where we define the function $z_{0}(m,n_r)$ for future reference. If this
result were exact, the quantized energies would depend on two quantum
numbers in at most a quadratic manner and there would be exact wave packet
revivals, just as for the 2D square or equilateral triangle billiards.

To see to what extent this approximation is, in fact, 
 valid for various values of
$m,n_r$, we evaluate (numerically)  a large number of the lowest-lying 
'exact' Bessel function zeroes, $z_{(m,n_r)}$. We then plot, in Fig.~1,  
the combination $2z_{(m,n_r)}/\pi - 3/2$ versus $m$ 
which would yield horizontal lines corresponding to constant, integral
values of $2n_r + |m|$ if the result in Eqn.~(\ref{first_approximation})
were  exact. For three such states, we also illustrate the corresponding
radial probability density and note that for increasing values of $|m|$,
there are fewer radial nodes so that the appropriate values of $z= kR$ used
in the boundary condition are smaller and the approximation used in
Eqn.~(\ref{first_approximation}) becomes worse.

\begin{figure}
\epsfig{file=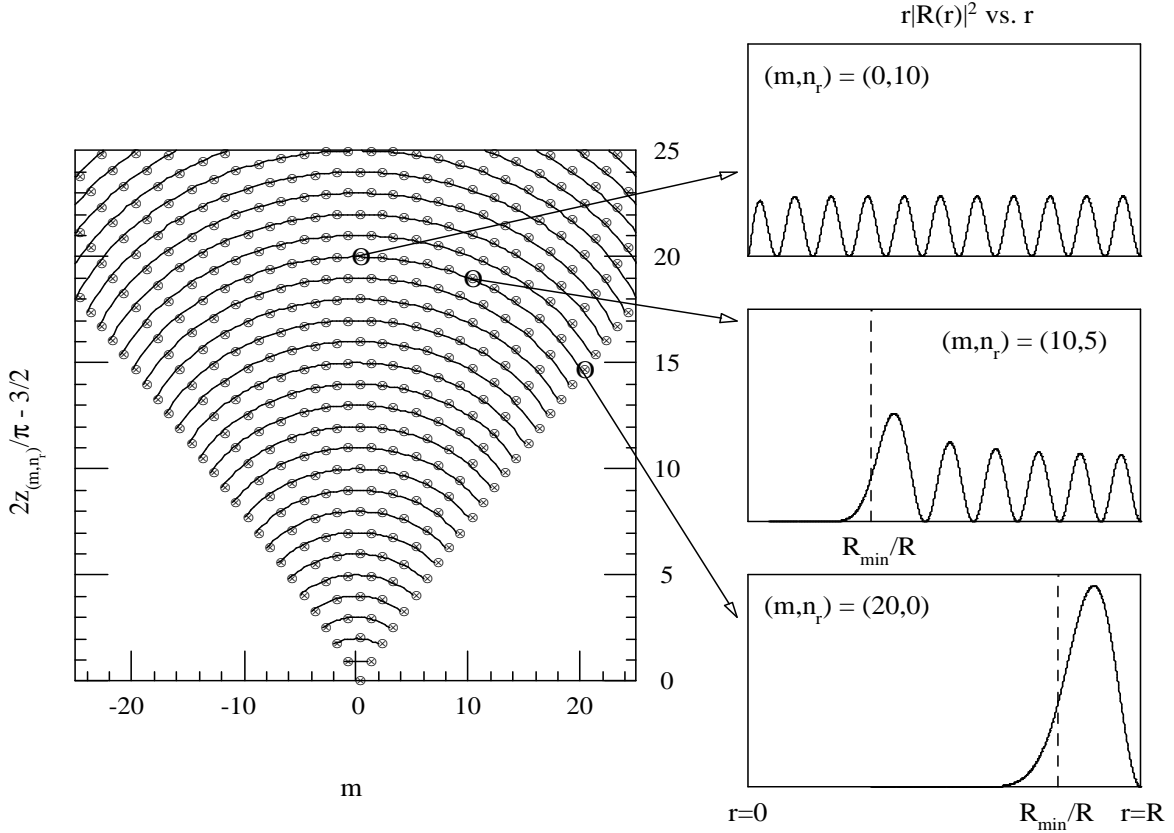,width=11cm,angle=270}
\caption{
Plot of the lowest-lying $z_{(m,n_r)}$ zeroes 
of the Bessel function, determined (numerically)  by 
$J_{|m|}(z_{(m,n_r)}) = 0$, 
versus $m$, scaled as $2z_{(m,n_r)}/\pi - 3/2$. Zeroes corresponding to 
fixed values of the combination $2n_r+|m|$ are shown connected, not
'fit' to data. Shown on the right are plots of the radial
probability density, $P(r) = r|R(r)|^2$ versus $r$ for the three states
shown with a 'circle' for which a value of $2n_r+|m|=20$ is fixed. The
value of $R_{min}/R = |m|/z$ suggested by Eqn.~(\ref{useful_form}) is
also indicated by a vertical dashed line.}
\end{figure}

 We see that the approximation is only a good one for $|m| \approx 0$ with 
obvious quadratic corrections. 
Instead of attempting to evaluate Bessel function zeros to higher
precision using more elaborate 'handbook' expansions of $J_{|m|}(z)$ 
for large $z$, we note that the  result of Eqn.~(\ref{first_approximation}), 
and the important and necessary corrections  to it, can be obtained in a 
straightforward and accessible way  by use of the WKB approximation. 

If we first quantize the angular variable to find that angular momentum is 
given by $L_z = m\hbar$, we can note that in the radial direction the particle 
moves freely up to the infinite wall at $r=R$, but is subject to an effective 
centrifugal
potential given by $V_{eff}(r) = L_z^2/2\mu r^2 = (m\hbar)^2/2\mu r^2$. 
The classical particle cannot penetrate this centrifugal barrier and has an 
associated inner radius (distance of closest approach) given by
\begin{equation}
V_{eff}(R_{min}) = \frac{m^2\hbar^2}{2\mu R_{min}^2} = E
\qquad
\mbox{or}
\qquad
R_{min} = \frac{|m|\hbar}{\sqrt{2\mu E}}
\end{equation}
We can also write this in the useful form
\begin{equation}
R_{min} = \frac{|m|R}{z}
\qquad
\mbox{where}
\qquad
E \equiv \frac{\hbar^2 z^2}{2\mu R^2}
\label{useful_form}
\end{equation}
more directly in terms of the desired dimensionless variable $z$,
which is equivalent to the energy eigenvalue. In Fig.~1, we have also 
indicated the value of $R_{min}/R = |m|/z$ for several of the radial 
probability densities, to compare to the fully quantum mechanical results.

The WKB quantization condition on the radial variable, $r$,
is then given by
\begin{equation}
\int_{R_{min}}^{R} k_r(r)\,dr = (n_r + C_L + C_R) \pi
\qquad
\mbox{where}
\qquad
k_r(r) \equiv  \sqrt{\frac{2\mu E}{\hbar^2}} \sqrt{1 - \frac{R_{min}^2}{r^2}}
\end{equation}
The matching coefficients \cite{wkb_approximation} are given by 
$C_L = 1/4$ and $C_R = 1/2$ which are appropriate for 'linear' boundaries 
(at the inner centrifugal barrier) and 'hard' or 'infinite wall' boundaries 
(such as at $r=R$), respectively. 
The WKB energy quantization condition for the quantized energies, 
in terms of $n_r$ explicitly and $|m|$ implicitly, via both the $E$ and 
$R_{min}$ terms, can then be written in the form
\begin{equation}
\sqrt{\frac{2\mu E}{\hbar^2}} 
\int_{R_{min}}^{R} \sqrt{1 - \frac{R_{min}^2}{r^2}} \,dr
= (n_r + 3/4)\pi
\label{wkb_condition}
\end{equation}
The integral on the left can be evaluated  in the form
\begin{eqnarray}
\int_{R_{min}}^{R} \sqrt{\frac{r^2 -R_{min}^2}{r}} \,dr
& = &
\sqrt{R^2 - R_{min}^2} - R_{min}\sec^{-1} \left(\frac{R}{R_{min}}\right) \\
& = & R \left[\sqrt{1-x^2} - x \sec^{-1}(1/x) \right] \nonumber
\end{eqnarray}
where we define $x \equiv R_{min}/R = |m|/z$. 
This result can be expanded for small values of $x$ (i.e., $R_{min}/R << 1$
or $|m|/z << 1$) to obtain
\begin{equation}
\sqrt{1-x^2} - x\sec^{-1}(1/x) = 1 - \frac{\pi x}{2} + \frac{x^2}{2}
+ \frac{x^4}{24} + \frac{x^6}{80} + \frac{5x^8}{896} + \cdots
\end{equation}
The WKB quantization condition in Eqn.~(\ref{wkb_condition})
can then be written, in terms of $z$,  in the form
\begin{equation}
z\left(1 - \frac{\pi}{2}\frac{|m|}{z} + \frac{m^2}{2z^2} + 
\frac{m^4}{24z^4} + \cdots \right)
= (n_r+3/4)\pi
\end{equation}
If we keep only the first two terms on the left-hand side, we find that
\begin{equation}
z \approx  (n_r +|m|/2+3/4)\pi \equiv  z_{0}(m,n_r)
\label{lowest_order}
\end{equation}
which is the lowest-order result obtained directly from the limiting form 
of the wavefunction. 

To improve on this result, we simply keep successively higher order terms, 
solving iteratively at each level of approximation using a lower-order result 
for $z$, and we find the much improved approximation
\begin{equation}
z_{(m,n_r)} = z_0(m,n_r) - \frac{m^2}{2z_0(m,n_r)}
- \frac{7}{24} \frac{m^4}{[z_0(m,n_r)]^3} - \frac{83}{240} \frac{m^6}{[z_0(m,n_r)]^5} - \frac{6949m^8}{13440[z_0(m,n_r)]^7} + \cdots
\label{zero_expansion}
\end{equation}
which we have confirmed numerically is an increasingly good approximation,
especially for $n_r >> 1$. For the study of wave packet revivals, we only
require the energy eigenvalue dependence on $m,n_r$ to second order, but
higher order terms such as those above might  be useful for super-revivals
and even longer-term time-dependence studies (or more detailed analytic
periodic orbit theory studies of the circular well.)

For the special case of $m=0$, we find no improvement using this WKB 
technique,  but motivated  by the form of the expansion in 
Eqn.~(\ref{zero_expansion}), we fit the first $50$ lowest-lying $m=0$
zeros to a similar form and find the result
\begin{equation}
z_{(0,n_r)} = z_0(0,n_r) + \frac{1}{8z_0(0,n_r)} - \frac{1}{24[z_0(0,n_r)]^3}
+ \cdots
\label{zero_zero_approximation}
\end{equation}
We cannot unambiguously fit to any higher-order terms, as much of the
non-linear spacing information is contained in the lowest few zeros.

Using Eqns.~(\ref{zero_expansion}) and (\ref{zero_zero_approximation}),
we can evaluate the energy eigenvalues to quadratic order in $n_r,m$ in
order to probe the revival structure of wave packets. For the special 
case of $m=0$, we find that
\begin{equation}
E_{(0,n_r)} = \frac{\hbar^2 [z_{(0,n_r)}]^2}{2\mu R^2}
= \frac{\hbar^2 \pi^2}{2\mu R^2}\left[\left(n_r+\frac{3}{4}\right)^2 + 
\frac{1}{4\pi^2}\right]
\label{zero_energies}
\end{equation}
while for the more general case with $m\neq 0$, we find
\begin{equation}
E_{(m,n_r)} = \frac{\hbar^2 [z_{(m,n_r)}]^2}{2\mu R^2}
= \frac{\hbar^2 \pi^2}{2\mu R^2}
\left[\left(n_r + \frac{|m|}{2} + \frac{3}{4}\right)^2 - \frac{m^2}{\pi^2}
\right]
\label{other_energies}
\end{equation}

The fact that these energies depend on non-integral values of the
effective quantum numbers is reminiscent of the case of Rydberg
wave packets in alkali-metal atoms due to quantum defects 
\cite{quantum_defects} and methods similar to those used there might
prove useful. In what follows, however, we simply examine the time-dependence 
of typical $m=0$ and $m \neq 0$ wave packets directly.

\section{ Gaussian wave packets and revivals}

Any wave packet in the circular billiard can be expanded in the 
normalized eigenstates of the form
\begin{equation}
\psi_{(m,n_r)}(r,\theta) = \left[N_{(m,n_r)}J_{|m|}(k_{(m,n_r)}r)\right]
 \left(\frac{1}{\sqrt{2\pi}}
e^{im\theta}\right)
\end{equation}
where
\begin{equation}
\left[N_{(m,n_r)}\right]^2 \int_{0}^{R} r\,\left[J_{|m|}(kr)\right]^2\,dr
= 1
\end{equation}
with expansion coefficients given by
\begin{equation}
a_{(m,n_r)} = \langle \psi(r,\theta;t=0) | \psi_{(m,n_r)} \rangle
\end{equation}
which satisfy
\begin{equation}
\sum_{m=-\infty}^{+\infty}\sum_{n_r=0}^{\infty}
|a_{(m,n_r)}|^2 = 1
\label{normalization}
\end{equation}
The expectation value of the energy in this potential well is given
by 
\begin{equation}
\langle \hat{E} \rangle 
= 
\sum_{m=-\infty}^{+\infty}\sum_{n_r=0}^{\infty}
|a_{(m,n_r)}|^2 \left(\frac{\hbar^2[z_{(m,n_r)}]^2}{2\mu R^2}\right)
\label{general_energy}
\end{equation}
and expectation values of powers of angular momentum are also easily
evaluated to give
\begin{equation}
\langle \hat{L}_z^{k} \rangle
= 
\sum_{m=-\infty}^{+\infty} 
\sum_{n_r = 0}^{\infty} |a_{(m,n_r)}|^2 (m\hbar)^{k}
\label{general_angular_momentum}
\end{equation}

The subsequent time-dependence of the wave packet is then given by
\begin{equation}
\psi(r,\theta;t) = \sum_{m=-\infty}^{+\infty} \sum_{n_r=0}^{\infty}
a_{(m,n_r)} \psi_{(m,n_r)}(r,\theta) \, e^{-iE_{(m,n_r)}t/\hbar}
\end{equation}
and the standard  autocorrelation function \cite{autocorrelation_function}
is given by 
\begin{equation}
A(t) \equiv 
\langle \psi(r,\theta;t) | \psi(r,\theta,0) \rangle
=
\sum_{m=-\infty}^{+\infty} \sum_{n_r=0}^{\infty} 
|a_{(m,n_r)}|^2  e^{-iE_{(m,n_r)}t/\hbar}
\label{autocorrelation_function}
\end{equation}

For definiteness, we will use a standard Gaussian wave packet of the
form 
\begin{equation}
\psi(x,y;t=0) = \psi_0(x;x_0,p_{0x},b)\psi_0(y;y_0,p_{0y},b)
\label{initial_gaussian}
\end{equation}
where
\begin{equation}
\psi_0(x;x_0,p_{0x},b) = \frac{1}{\sqrt{b\sqrt{\pi}}} e^{ip_{0x}(x-x_0)/\hbar}
e^{-(x-x_0)^2/2b^2}
\end{equation}
with a similar expression for $\psi_0(y;y_0,p_{0y},b)$.
The initial expectation values for the $x$ variables are given by
\begin{equation}
\langle x\rangle_0 = x_0
\qquad
,
\qquad
\langle x^2 \rangle_0 = x_0^2 + \frac{b^2}{2}
\qquad
,
\qquad
\Delta x_0 = \frac{b}{\sqrt{2}}
\end{equation}
and
\begin{equation}
\langle p_x\rangle_0 = p_{0x}
\qquad
,
\qquad
\langle p^2_x \rangle_0 = p_{0x}^2 + \frac{\hbar^2}{2b^2}
\qquad
,
\qquad
\Delta p_0 = \frac{\hbar }{\sqrt{2}b}
\end{equation}
with similar results for $y$. So long as the initial location, 
$(x_0,y_0)$, is well away from the edges of the potential well, such a
Gaussian form can be easily and reproducibly expanded in terms of eigenstates. 
The expectation value of total energy is 
\begin{equation}
\langle \hat{E} \rangle = \frac{1}{2m} \langle \hat{p}_x^2 
+ \hat{p}_y^2 \rangle
= \frac{1}{2m} \left[(p_{0x})^2 + (p_{0y})^2 + \frac{\hbar^2}{b^2}\right]
\label{gaussian_energy}
\end{equation}

In this central potential, angular momentum is conserved and we also
have the specific results for this Gaussian form
\begin{equation}
\langle \hat{L}_z \rangle
=
\langle x \hat{p}_y - y \hat{p}_x \rangle
= \langle x \rangle \langle \hat{p}_y \rangle
- \langle y \rangle \langle \hat{p}_x \rangle
= x_0p_{0y} - y_0 p_{0x}
\label{gaussian_angular_momentum_1}
\end{equation}
and
\begin{equation}
\langle \hat{L}_z^2 \rangle
= (x_0p_{0y} - y_0 p_{0x})^2
+ \frac{b^2}{2} \left[(p_{0x})^2 + (p_{0y})^2\right]
+ \frac{\hbar^2}{2b^2} \left[(x_0)^2 + (y_0)^2\right]
\label{gaussian_angular_momentum_2}
\end{equation}
so that 
\begin{equation}
(\Delta m )\hbar 
\equiv 
\Delta L_z = 
\sqrt{
\frac{b^2}{2} \left[(p_{0x})^2 + (p_{0y})^2\right]
+ \frac{\hbar^2}{2b^2} \left[(x_0)^2 + (y_0)^2\right]
}
\label{gaussian_angular_momentum_spread}
\end{equation}

As a check on the numerical evaluation of the expansion coefficients, 
it is useful to be able to compare the general results for 
$\langle E \rangle$ and $\langle \hat{L}_z^{(1,2)} \rangle$ in 
Eqns.~(\ref{general_energy}) and (\ref{general_angular_momentum}) 
with the specific results for the Gaussian in Eqns.~(\ref{gaussian_energy}), 
(\ref{gaussian_angular_momentum_1}), and (\ref{gaussian_angular_momentum_2}).

We begin by focusing on the special case of zero-momentum wave packets
centered at the origin, namely with vanishing values of 
$(p_{0x},p_{0y})$ and $(x_0,y_0)$ in which case the initial wave packet
is radially symmetric and therefore has an expansion in pure $m=0$
angular momentum states. (This is consistent with the result in
Eqn.~(\ref{gaussian_angular_momentum_spread}) which has $\Delta L = 
\Delta m = 0 $ for this state).

For such states, where only the $m=0$ eigenstates contribute, we can write
the energy eigenvalues from Eqn.~(\ref{zero_energies}) in the form
\begin{eqnarray}
E(n_r) & = & \frac{\hbar^2\pi^2}{2\mu R^2}
\left[ \left(n_r+\frac{3}{4}\right)^2 + \frac{1}{4\pi^2}
+ {\cal O}\left(\frac{1}{(n_r+3/4)^2}\right) \right]
\nonumber  \\
& \approx  & 
\frac{\hbar^2\pi^2}{32\mu R^2}\left[ 8n_r(2n_r+3) + 
\left(9 + \frac{4}{\pi^2}\right)\right] 
\label{constant_piece}\\
& = & \frac{\hbar^2 \pi^2}{4\mu R^2} 
\left[ l(n_r) + \left(\frac{9}{8} + \frac{1}{2\pi^2}\right)\right] \nonumber
\end{eqnarray}
where $l(n_r) \equiv n_r(2n_r+3)$ is an integer 
(neither even nor odd in general).
The last term in the square brackets is independent of $n_r$ and will make
the same, constant, overall phase contribution to the autocorrelation
function,  so we focus on the $l(n_r)$ term. Since this integer has no
special evenness/oddness properties, its contribution to the phase of
each $|a_{(n,n_r)}|^2$ term in Eqn.~(\ref{autocorrelation_function})
will be identically unity at a revival time given by
\begin{equation}
\left(\frac{\hbar^2 \pi^2}{4\mu R^2}\right) \frac{T_{rev}^{(m=0)}}{\hbar}
=  2\pi
\qquad
\mbox{or}
\qquad
T_{rev}^{(m=0)} = 4\left[\frac{2\mu R^2}{\hbar \pi}\right] \equiv
4T_0
\end{equation}
Thus, at integral multiples of $4T_0$, we expect nearly perfect revivals
because of the almost regularly spaced structure of the $m=0$ Bessel function
zeros. At these recurrences, we can also predict  the overall phase
corresponding to the last term in Eqn.~(\ref{constant_piece}), namely
\begin{equation}
e^{-i\hbar^2\pi^2/4\mu R^2(4T_0) (9/8+1/2\pi^2)}
= e^{-2\pi i (9/8 +1/2\pi^2)}
=
e^{-2\pi i} e^{-2\pi i(1/8+1/2\pi^2)}
\equiv e^{-i\pi F}
\label{extra_phase_1}
\end{equation}
where $F = 1/4 + 1/\pi^2 \approx 0.351$. 

To investigate these predictions numerically, we have used a Gaussian of the 
form in Eqn.~(\ref{initial_gaussian}) with the specific  values
\begin{equation}
2m = \hbar = R = 1
\qquad
\mbox{and}
\qquad
b = \frac{1}{10\sqrt{2}}
\qquad
\mbox{so that}
\qquad
\Delta x_0 = \Delta y_0 = 0.05
\label{numerical_values}
\end{equation}
Using the normalized eigenstates, we numerically evaluate the overlap
integrals to obtain the $a_{(m,n_r)}$, using enough states to ensure
that the appropriate conditions, such as Eqns.~(\ref{normalization}),
(\ref{general_energy}), and (\ref{general_angular_momentum}),
are all satisfied to better than $10^{-4}$ accuracy. 

Using the expansion coefficients for this state, we plot the modulus squared
of the autocorrelation function, $|A(t)|^2$, in the bottom plots of both
Figs.~2 and 3, with time 'measured' in units of $T_0$. The almost exact
revival structure at integral multiples of $4T_0$ is evident. As a further
check, we can evaluate the phase of $A(t)$ at each revival and find that
to an excellent approximation it is given by $-nF\pi$ as in 
Eqn.~(\ref{extra_phase_1}). 
If one decreases (increases) the value of $b$,
 so that the initial wave packet is narrower (wider), 
the energy eigenvalues  required to construct the packet are
then larger (smaller) (from Eqn.~(\ref{gaussian_energy})) and 
are therefore generally more (less) evenly spaced 
(from Eqn.~(\ref{zero_zero_approximation})) and we indeed confirm 
this with our numerical simulations; 
the eventual, long-term  decrease $|A(nT_0)|$ with increasing $n$ is 
faster (slower) for smaller (larger) values of $b$.

\begin{figure}
\epsfig{file=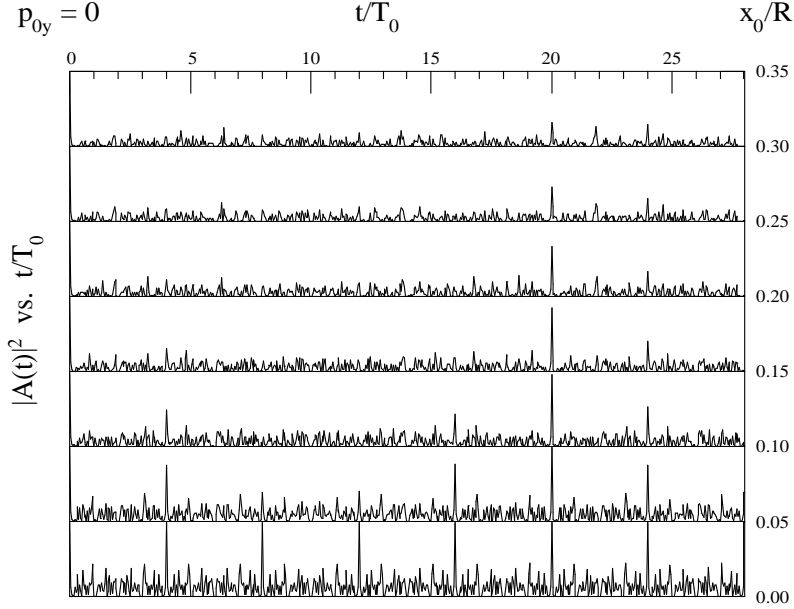,width=8cm,angle=270}
\caption{
Plot of the autocorrelation function, $|A(t)|^2$
versus $t$, in units of $T_0 \equiv  2\mu R^2/\hbar \pi$, The numerical values
of Eqn.~(\ref{numerical_values}) are used, along with $y_0= 0$ 
and $p_{0x}=p_{0y} = 0$. The results for $|A(t)|^2$ versus $t$, 
 as one varies the $x_0$ of the  initial wave packet away from the
center of the circular billiard, are shown on horizontal axes.
}
\end{figure}

\begin{figure}
\epsfig{file=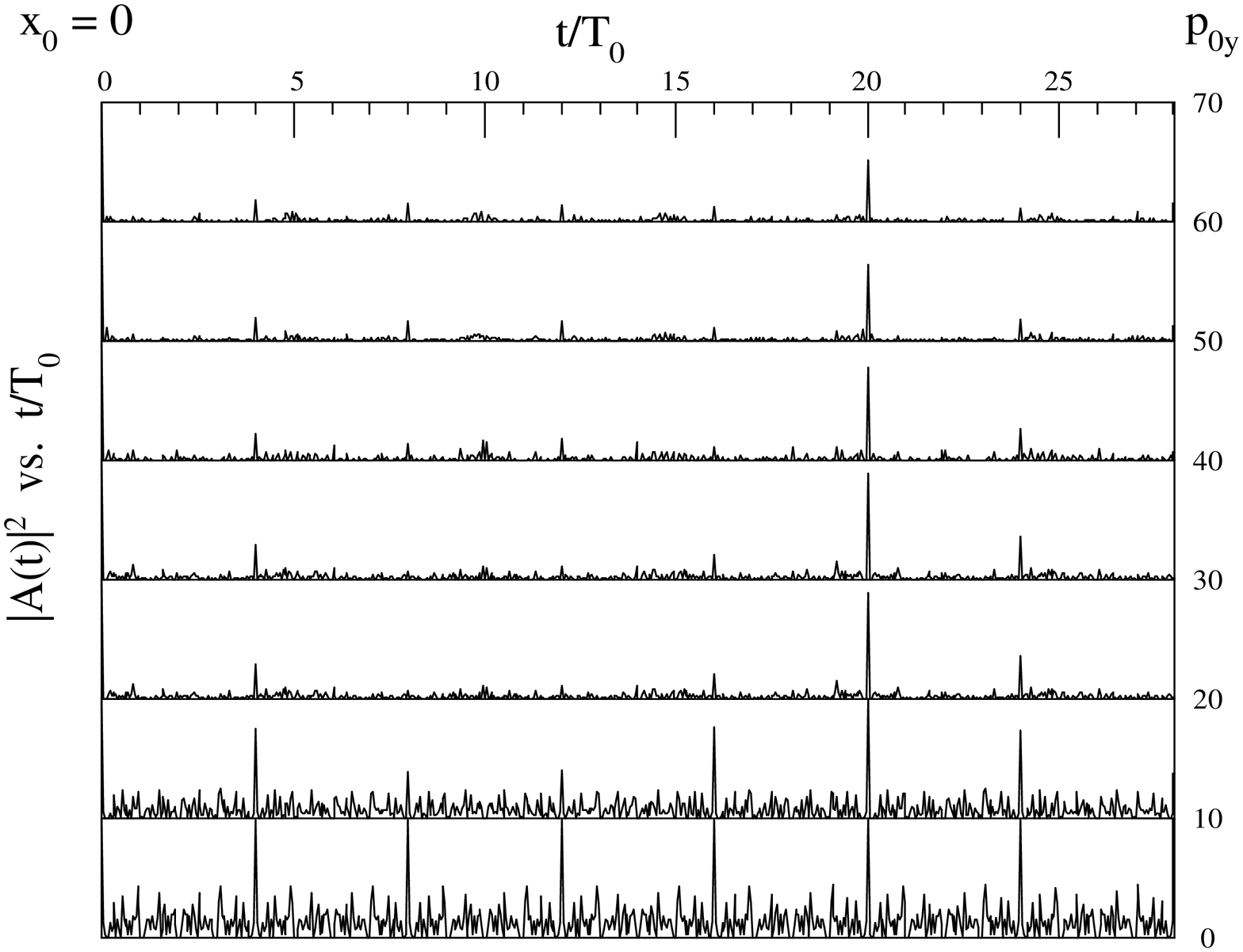,width=8cm,angle=270}
\caption{
Same as Fig.~2, but with $x_0=y_0=0$ and
$p_{0x}=0$, as one increases $p_{0y}$.
}
\end{figure}

We next move away from the special case of the zero-momentum, central
wave packet by considering individually the case of $x_0 \neq 0$ and
 $p_{0y} \neq 0$ (but not both). In each case,  the average angular momentum
of the state is still vanishing 
(from Eqn.~(\ref{gaussian_angular_momentum_1})), 
but $m\neq 0$ values of the expansion coefficients are now required.
We must now use the more general case for the energies, which to 
second order in $m\neq 0, n_r$, are given by Eqn.~({\ref{other_energies})
\begin{eqnarray}
E_{(m,n_r)} & = & \frac{\hbar^2 \pi^2}{2\mu R^2} 
\left[ \left(n_r + \frac{|m|}{2} + \frac{3}{4}\right)^2 
- \frac{m^2}{\pi^2} \right] \nonumber \\
& = & \frac{\hbar^2 \pi^2}{32 \mu R^2}
\left[ (16n_r^2 + 24n_r + 16|m|n_r) + 4|m|(|m|+3) - \frac{16 m^2}{\pi^2}
\right] \\
\label{approximate_energies}
& = & \frac{\hbar^2 \pi^2}{32 \mu R^2}
\left[ 8\tilde{l}(n_r) + 8\overline{l}(|m|) - \frac{16 m^2}{\pi^2} +9 \right]
\nonumber
\end{eqnarray}
where 
\begin{equation}
\tilde{l}(n_r) \equiv n_r(2n_r +3 +2|m|)
\qquad
\mbox{and}
\qquad
\overline{l}(|m|) \equiv |m|(|m|+3)/2
\end{equation}
are both integers, again, with no special even or oddness properties.
We can then write these energies in the form
\begin{equation}
E_{(m,n_r)} = \frac{2\pi \hbar}{4T_0}\left[\tilde{l}(n_r) + \overline{l}(|m|)
- \frac{2m^2}{\pi^2} + \frac{9}{8} \right]
\end{equation}
At integral multiples of the $m=0$ revival time, $t_N = N(4T_0)$, 
the first two terms give $e^{-N(2\pi i)} = 1$ phases to each $(m,n_r)$ 
term in the autocorrelation function, while the last term gives an overall, 
$(m_,n_r)$-independent  phase,  just as in the $m=0$ case.  
The other term, however,  gives a contribution
\begin{equation}
e^{-(2\pi i) (m^2 N) (2/\pi^2)}
\end{equation}
which depends on $m$ explicitly and which therefore eliminates the
revivals, increasingly so, as the wave packet is dominated by $m \neq 0$
terms. However, because of a seeming numerical accident, at integral
multiples of $5T^{(m=0)}_{rev} = 20 T_0$, we recover approximate revivals
due to the fact that $5 \times (2/\pi^2) = 1.013$. We thus find approximate
revivals for the more general $m \neq 0$ case given by
$T_{rev}^{(m \neq 0)} = (\pi^2/2) T_{rev}^{(m=0)} \approx 5T_{rev}^{(m=(0))}$.

This effect is illustrated in more detail in Figs.~2 and 3 where we
plot $|A(t)|^2$ versus $t$ as we move from the central, zero-momentum
wave packet by first moving away from the origin ($x_0 \neq 0$ in Fig.~2)
or having non-zero momentum values ($p_{0y} \neq 0$ in Fig.~3).
In each case, as we increase the parameter ($x_0$ or $p_{0y}$),
 we necessarily
include more and more $|m| \neq 0$ eigenstates. For even a small mix of
such states, the $T_{rev}^{(m=0)}$ revival periods at most integral
multiples of $4T_0$ disappear, while evidence for the more general
$T_{rev}^{(m \neq 0)} = 20 T_{0}$ revivals remains evident.

For the particular numerical values used in Eqn.~(\ref{numerical_values}),
the spread in angular momentum required from
Eqn.~(\ref{gaussian_angular_momentum_spread}) is given by
\begin{equation}
\Delta L = \sqrt{\left(\frac{p_{0y}}{20}\right)^2 + (10x_0)^2}
= \Delta m
\end{equation}
(since $\hbar = 1$) so that the $x\neq 0$ and $p_{0y} \neq 0$ values 
used in Figs.~2 and 3 actually correspond to the same set of $\Delta L$ 
for each horizontal case shown.

We note that this 'lifting' of a seemingly 'accidental' degeneracy in the 
pattern of revival times is somewhat similar to the special case of a 
zero-momentum Gaussian wave packet in a 2D square or triangular 
billiard, initially placed at the center, cases which we briefly discuss 
in Appendix A.

This pattern of revival times depending on two distinct quantum numbers
is also somewhat reminiscent of that encountered in a rectangular billiard
with differing sides of length $L_x,L_y$ where if the sides are
incommensurate one would expect a less elaborate revival structure. Since the
revival times typically scale as $T_{rev} \propto L^2$, the appearance of a
$\pi^2$ scale factor which can  give rise to very close to  an integer ratio 
$10/\pi^2 \approx 1$ (to within $1.3\%$) is appropriate; in this case, 
the  relevant length scales for the radial quantum number and azimuthal
quantum numbers are most likely multiples of $R$ and  $2\pi R$ respectively,
so that relative factors of $\pi^2$ in the revival times can appear naturally. 

The presence of the $\Delta m \neq 0$ revivals becomes increasingly
less obvious as the average angular momentum is increased away from zero
(with both $x_0$ and $p_{0y}$ now non-vanishing), 
since the  required energy eigenvalues are in a region of large
$|m|/z$ where the lowest-order approximation (from Eqn.~(\ref{lowest_order}))
of evenly spaced $z$ values becomes worse. We also note that even with
$\langle \hat{L} \rangle = 0 $, as we increase  $p_{0y}$, the spread
in $m$ values required also increases 
(as in Eqn.~(\ref{gaussian_angular_momentum_spread})),
so that the overall number of states required to reproduce the initial
Gaussian, and which have to 'beat' against each other appropriately, 
increases as well,
making revivals more difficult to produce. The increasingly large number
of states required to construct the Gaussian wave packets for larger values
of $\langle \hat{L} \rangle$ can also be seen during the collapsed phase when
the 'average value' of $|A(t)|^2$, namely $\sum_{m,n_r} |a_{(m,n_r)}|^4$,
becomes increasingly small as the fixed probability 
(constrained via $\sum_{m,n_r} |a_{(m,n_r)}|^2$ = 1) 
is spread over more and more states.

While we have focused on the long-term, revival structure of the
autocorrelation function, the appearance of more short-time 
 features in $|A(t)|$, corresponding to short-term, semi-classical 
closed orbits, is also apparent. Such trajectories are characterized
\cite{balian_and_bloch} by periodic  orbits with path lengths and minimum radii
(distances of closest approach) given by
\begin{equation}
L(p,q) = 2p R \sin\left(\frac{\pi q}{p}\right)
\qquad
\mbox{and}
\qquad
R_{min} = R \cos\left(\frac{\pi q}{p}\right)
\label{classical_conditions}
\end{equation}
with integral values of $(p,q)$ (with $p>2q$) describing the number of 'hits' 
on the wall and the number of 'revolutions' for one complete orbit 
respectively. If, for example, we place Gaussian wave packets with
$p_{0x}=0$ and $p_{0y} > 0$ at locations given by 
$(x_0,y_0) = (R_{min}(p,q),0)$, we find obvious peaks in the autocorrelation 
function at times given by $T_{cl}(p,q) \equiv L(p,q)/(p_{0y}/m)$ 
corresponding to classical closed orbits. (This structure is evident, of
course, only when the expected classical periods are less than the wave 
packet spreading time, $\Delta t$; this can be estimated using the result 
for a free Gaussian as $\Delta t = 2m\Delta x_0^2/\hbar$.) We discuss in 
Appendix B exactly how the classical closed orbit periodicity is reproduced 
from the quantum mechanical energy spectrum, using the WKB approximation of 
Eqn.~(\ref{wkb_condition}).

Variations on the problem of a circular infinite well can  also be examined
for their possible revival structure. The 'half-circular' well, with an
infinite wall added along a diameter,  is exactly soluble with linear 
combinations of $e^{im\theta}$ and $e^{-im\theta}$ solutions being able to
satisfy the new boundary condition for $m \neq 0$, while the $m=0$ solutions
are not allowed.  The energy spectrum then consists of one copy
 of the $E_{(m\neq 0,n_r)}$ values for the circular well, with a resulting
revival behavior consistent with $T_{rev}^{(m\neq 0)}$,
 since no $m=0$ states are allowed. 

Another variant would be an annular circular billiard, with an inner
infinite wall at $r=R_{inner} < R$. The energy eigenstates can also be 
derived using Bessel function solutions (now including the 'irregular' or
divergent $Y_{|m|}(kr)$ terms since the particle is kept explicitly away 
from the origin by the inner wall) with the energy eigenvalues resulting from 
the condition
\begin{equation}
J_{|m|}(kR)
Y_{|m|}(kR_{inner})
-
J_{|m|}(kR_{inner})
Y_{|m|}(kR)
= 0
\end{equation}
WKB-type expansions for the quantized energies are also useful in this
case.  The qualitatively new features
present in this geometry include new classical orbits (bouncing off the
inner wall) but also diffraction features, as seen in periodic orbit
theory analyses \cite{annular_billiard} of such systems.  (For the WKB
analysis of the energies corresponding to orbits which bounce off the
'hard' inner wall, one must use $(n_r+1)$ in place of $(n_r+3/4)$ in 
Eqn.~(\ref{wkb_condition}).) While
$m=0$ states are allowed, the special central, zero-momentum initial state
is not, and whether a pattern of something like the 
$T_{rev}^{(m\neq 0)}$ revivals is supported is currently
under study.

\begin{acknowledgments}
This work was supported in part by the National Science Foundation
under Grant DUE-9950702. 
\end{acknowledgments}

\appendix

\section{}

We briefly consider the special case of a Gaussian wave packet 
(of the form in Eqn.~(\ref{initial_gaussian})) with
vanishing momentum and initially located at the center of a two-dimensional
square infinite well (or billiard) of dimension $L \times L$. Because
the problem is entirely separable, the autocorrelation function for the
2D problem will be a product of the individual 1D values, namely
$A(t) = A_{x}(t) \times A_{y}(t)$, so it suffices to consider the 1D case.
The energy eigenstates and eigenvalues are given 
\begin{equation}
u_n(x) = \frac{2}{L} \sin\left( \frac{n\pi x}{L} \right)
\qquad \mbox{and} \qquad
E_n = \frac{\hbar^2 \pi^2 n^2}{2mL^2}
\qquad
\mbox{with $n=1,2,3,4,...$}
\end{equation}
(over the range $(0,L)$) 
and the general revival time is given by $T_{rev} = 4mL^2/\hbar \pi$.
For an initial Gaussian with vanishing momentum ($p_{0x} = 0$) and
located at the center of the well ($x_0=L/2$), 
the 1D expansion coefficients, $a_{n}$,  simplify
since the 'odd' parity  states (here meaning $n=2,4,...$) make no contributions
to the wave packet and the energies can be written in the form
\begin{equation}
E_{n} = \frac{\hbar \pi^2}{2mL^2} (2n-1)^2 
\qquad
\mbox{with $n = 1,2,3,...$}
\end{equation}
or as
\begin{equation}
E_{n} = \frac{\hbar^2 \pi^2}{2mL^2} 
\left[ 4n^2 - 4n +1\right]
= 
\frac{\hbar^2 \pi^2}{2mL^2} 
\left[ 8n(n-1)/2 + 1 \right]
= 
\left(\frac{2\pi \hbar}{T_{rev}}\right) 8 \left[\tilde{N}(n) + 1/8\right]
\end{equation}
where $\tilde{N}(n) \equiv n(n-1)/2$ is an integer (neither even nor odd
in general). Just as with the  $m=0$  case of the circular well,
in this very special alignment, the modulus of $A(t)$ is unity with a
reduced revival time of $T_{rev}^{(center)} = T_{rev}/8$, with a 
predictable phase factor (due to the  constant, $1/8$ term) 
at integral multiples of $T_{rev}^{(center)}$. If one moves away from 
this special
case by having $x_0 \neq L/2$ or $p_0 \neq 0$, this special revival
structure is lost and only the (still exact) $T_{rev}$ revivals are
evident.

\begin{figure}[hbt]
\epsfig{file=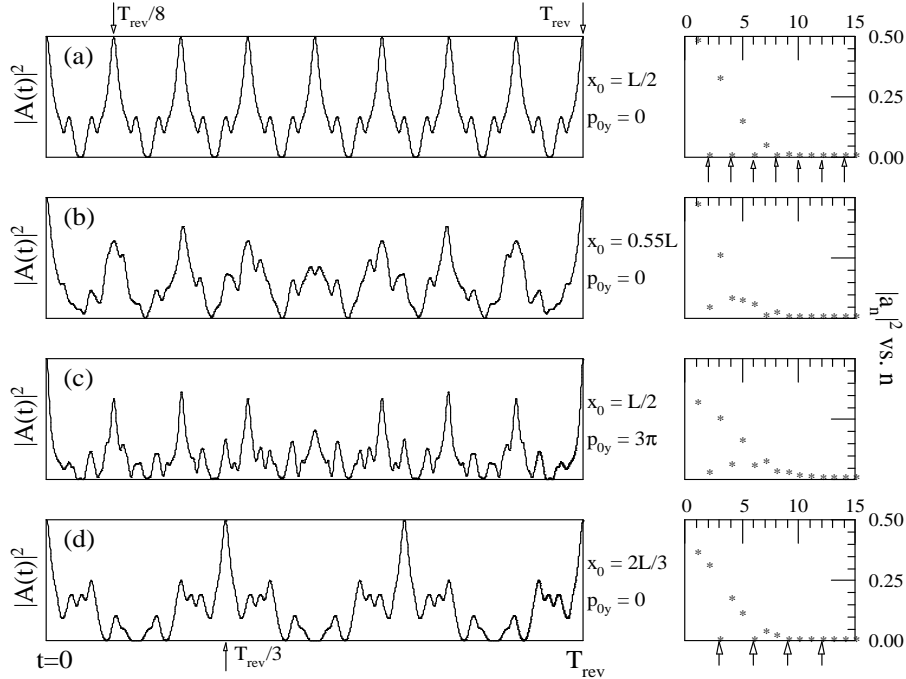,width=9cm,angle=270}
\caption{Plots of the one-dimensional autocorrelation
function, $A(t)$ versus $t$, for various Gaussian wave packets
in a one-dimensional infinite well defined over the length $(0,L)$.
For case (a),  where $x_0=L/2$ and $p_0 = 0$, there are special revivals
at $T_{rev}/8$ due to the extra symmetries forcing all even  expansion
coefficients to vanish. The values of $|a_n|$ are shown directly to the
right, with arrows indicating those which vanish identically for 
symmetry reasons. Cases (b) and (c) respectively show the effect of 
changing $x_0$ and $p_0$ slightly away from the values in case (a), 
illustrating how only the 'true' revival time is maintained. In case (d), 
we show another special case ($x_0 = L/3$) where certain $a_n$ vanish 
(every third one in this case) for symmetry reasons, also giving 
accidental revival times.}
\end{figure}

We illustrate this in Fig.~4 for the case of a
central, no-momentum solution (a) and for $x_0 \neq L/2$ (b) and
$p_0 \geqnew  0$ (c) cases. We also show in the bottom (d) of Fig.~4 another 
special
case where certain expansion coefficients vanish for symmetry reasons
in the no-momentum case ($x_0 = 2L/3$, where every third $a_n$ is
zero) with exact revivals at integral multiples of $T_{rev}/3$. Thus,
a no-momentum  2D Gaussian wave packet moved slightly away from say 
$(x_0,y_0) = (2L/3, L/3)$ would experience the same kind of 'broken' 
revival time symmetry, as one moved from the center. 

A similar set of 'accidental' or 'symmetry' revival times exists for
the equilateral triangle billiard. For example, for a zero-momentum
state placed at the geometrical center, the revival time is $T_{rev}/9$
where $T_{rev}$ is the exact revival time 
in Eqn.~(\ref{triangle_revival_time}).  Similar
'symmetry' points exist at distances of $\sqrt{3}L/12$ from the center
in the direction of each vertex where the revival times are $T_{rev}/4$.
Not surprisingly, we find no additional  such 'symmetry'  points 
(besides the center) in the circular case.

\newpage

\section{}

While we have focused on the longer-term, revival dependence of wave
packets in the circular well, it is interesting to note how the
information about the classical closed (or periodic) orbits in this
system is encoded in the energy eigenvalue spectrum, especially since most of
the experimentally observed 2D circular billiard systems 
\cite{microstructures}, \cite{atom_optics} have made measurements
which are  relevant for short-term, quasi-classical ballistic propagation. 
Such closed orbits are also the ones of relevance to periodic orbit
theory \cite{microwave} measurements of such billiard systems.

For a system with two quantum numbers, there are two classical periods 
\cite{bluhm_2d}, which in our case are given by
\begin{equation}
T_{cl}^{(n_r)} \equiv \frac{2\pi \hbar}{\partial E/\partial n_r}
\qquad
\mbox{and}
\qquad
T_{cl}^{(m)} \equiv \frac{2\pi \hbar}{\partial E/\partial m}
\end{equation}
and the two periods can beat against each other to produce the
classical periodicity ($T_{cl}^{po}$) for closed or periodic orbits 
if they satisfy 
\begin{equation}
pT_{cl}^{(n_r)} = T_{cl}^{(po)} = q T_{cl}^{(m)}
\end{equation}
with $p>2q$ for this geometry. We can then use this formalism to 
understand how these conditions can
give rise to the classical expressions for the minimum radius and path
lengths in Eqn.~(\ref{classical_conditions}). Instead of using the approximate
expression in Eqn.~(\ref{approximate_energies}) for the $(m,n_r)$-dependent
energies, we make use of the WKB condition in Eqn.~(\ref{wkb_condition})
and simply take partial derivatives of both sides with respect to
$n_r$ and $m$ respectively. We thus obtain the conditions
\begin{eqnarray}
\sqrt{\frac{\mu}{2\hbar^2}}
\left[ \int_{R_{min}}^{R} \frac{dr}{\sqrt{E - m^2\hbar^2/2\mu r^2}} \right]
\left(\frac{\partial E}{\partial n_r} \right) & = & \pi \\
\sqrt{\frac{\mu}{2\hbar^2}}
\left[ \int_{R_{min}}^{R} \frac{dr}{\sqrt{E - m^2\hbar^2/2\mu r^2}} 
\left(\frac{\partial E}{\partial n_r} 
- \frac{|m|\hbar^2}{\mu r^2}\right) \right]
& = & 0
\end{eqnarray}
The condition to be satisfied for periodic orbits can then be written
as
\begin{equation}
\frac{q}{p} = \frac{T_{cl}^{(n_r)}}{T_{cl}^{(m)}}
= \frac{(\partial E/ \partial m)}{(\partial E/\partial n_r)}
= \left(\frac{|m|\hbar}{\pi \sqrt{2\mu E}} \right)\left[\int_{R_{min}}^{R}
\frac{dr}{r\sqrt{r^2 - R_{min}^2}}\right]
\end{equation}
Evaluating the integral and using $R_{min} \equiv |m|\hbar/\sqrt{2\mu E}$,
we find that
\begin{equation}
\frac{q}{p} =\frac{1}{\pi} \sec^{-1}\left(\frac{R}{R_{min}}\right)
\qquad
\mbox{or}
\qquad
R_{min}(p,q) \equiv R_{min} = R \cos\left(\frac{\pi q}{p}\right)
\end{equation}
as the condition on periodic orbits, as expected. 
 To find the classical period for such closed orbits, we note that
\begin{equation}
T_{cl}^{(po)} = p T_{cl}^{(n_r)} = 
 \frac{2\pi \hbar p }{(\partial E/\partial n_r)}
= \left(2p \sqrt{R^2 - R_{min}^2} \right) \sqrt{\frac{\mu}{2E}}
= \frac{[2pR\sin(\pi q/p)]}{v_0} = \frac{L(p,q)}{v_0}
\end{equation}
where we identify $v_0 = \sqrt{2E/\mu}$ with the classical speed.

The classical periods for the closed orbits for the 2D annular
well mentioned above  can also be obtained from the WKB approximation in 
the same way, including those for the new features corresponding to
'bounces' off the inner infinite wall when $R_{min}$ is replaced by
$R_{inner}$ in the integrations. The classical periods for the closed 
orbits for the 2D square and 
equilateral triangle billiards can, of course,  also be obtained 
in the identical manner,  using the exact results for their energies.

\end{document}